# Replacing discontinued Big Tech mobility reports: a penetration-based analysis

Francesco Finazzi[1,*]

[1]University of Bergamo, Department of Economics; Bergamo, Italy.
*Corresponding author. Email: francesco.finazzi@unibg.it

## Abstract

People mobility data sets played a role during the COVID-19 pandemic in assessing the impact of lockdown measures and correlating mobility with pandemic trends. Two global data sets were Apple's Mobility Trends Reports and Google's Community Mobility Reports. The former is no longer available, while the latter will be discontinued in October 2022. Thus, new products will be required.

To establish a lower bound on data set penetration guaranteeing high adherence between new products and the Big Tech products, an independent mobility data set based on 3.8 million smartphone trajectories is analysed to compare its information content with that of the Google data set. This lower bound is determined to be between $10^{-4}$ and $10^{-3}$ (1 trajectory every 10,000 and 1000 people, respectively) suggesting that relatively small data sets are suitable for replacing Big Tech reports.

## Summary

People mobility data received considerable attention during the COVID-19 pandemic when they were used to assess the impact of lockdown measures[1,2,3,4,5,6,7] and to understand the correlation between mobility patterns and pandemic trends[8,9,10,11,12,13,14]. The importance of mobility data is not limited to the pandemic. In general, mobility is affected by global long-term events such as economic crises[15,16,17] and conflicts[18] and by local short-term events like social unrest and extreme natural events[19,20,21,22]. Mobility data are also used to better assess exposure to health-threatening phenomena[23].

Smartphone mobility data play a key role in estimating mobility patterns[24,25,26,27]. Apple's Mobility Trends Reports and Google's Community Mobility Reports were two global data sets made available to researchers during COVID-19 pandemic[28,29,30,31,32]. The first data set was discontinued on April 14, 2022[33], while the second is expected to be discontinued in October 2022[34].

Here, a global mobility data set provided by a private company is analysed to assess the feasibility of making available to the scientific community mobility products which, in their information content, are similar to Apple and Google products. The analysed data set includes 6.1 billion location data points collected by smartphone apps from March 11, 2020, to September 22, 2022. From the data set, 3.8 million anonymized spatio-temporal smartphone trajectories are reconstructed and used to produce mobility metrics at the country level.

Similarly to Gao et al[3], two metrics are provided: daily average travel distance ($M_1$) and the percentage of people who did not move during the 24 hours of the day ($M_2$). Peculiar aspects of smartphone-based location data are addressed, such as: the non-negligeable uncertainty on smartphone coordinates, missing data and the non-homogeneous geographical penetration of smartphone apps in the population. These aspects are considered to estimate mobility metrics characterized by the lowest possible bias and accompanied by a measure of uncertainty.

Analysis is restricted to a group of 17 countries for which uncertainty on the estimated mobility metrics is small enough to allow reasonable comparisons between countries and/or different periods: Argentina (ARG), Chile (CHL), Colombia (COL), Costa Rica (CRI), Ecuador (ECU), Greece (GRC), Guatemala (GTM), Italy (ITA), Mexico (MEX), Nicaragua (NIC), Panama (PAN), Peru (PER), Philippines (PHL), Slovenia (SVN), Turkey (TUR), the United States (USA) and Venezuela (VEN).

Metrics $M_1$ and $M_2$ are provided at daily temporal resolution and at different levels of temporal smoothing, uncertainty included. Mobility metrics time series are correlated with time series of Google's product in order to compare their mutual information content and to assess a lower bound on the number of smartphone trajectories (with respect to the country population size) which guarantees a high adherence between products.

## Results

### Global long-term mobility trends

Mobility metrics and their smoothed versions are computed for the aforementioned countries. Fig. 1 and Fig. 2 show polar plots of $M_1$ and $M_2$ time series based on a 14-day temporal smoothing. All countries show significant decrements in the $M_1$ metric and significant increments in the $M_2$ metric during the few months after March 11, 2020 (initial phase of the COVID-19 pandemic). Differences between countries are observed in the temporal rapidity of the subsequent "recovery". Some exhibit "lobed" polar plots. This is the case for GRC, ITA and TUR, which show fast recoveries during the summer of 2020 and a contraction of people mobility during the subsequent winter. All other countries exhibit "spiralling" polar plots, which is a sign of a slow recovery. This behaviour is clear for South American countries like ARG, COL, PER and VEN.

A year-by-year comparison is made by considering the period from March 11 to September 22, which is covered by the data set in years 2020, 2021 and 2022. Daily (non-smoothed) figures of $M_1$ and $M_2$ are averaged over the above period for each year. Fig. 3 depicts the 3-year trends of the averaged $M_1$ and $M_2$ metrics for the 17 countries. Most countries exhibit a linear or near-linear trend in the relationship between $M_1$ and $M_2$, a sign that mobility patterns continuously evolved over the 3 years. The rate of this evolution, however, is country-specific. From 2021 to 2022, ARG and USA exhibit the largest change in the daily average distance travelled. NIC shows little variability over the 3 years, as though its people mobility was only mildly affected by the COVID-19 pandemic.

### Short-local events in mobility metrics

$M_1$ and $M_2$ time series are also affected by short-term local events. Fig. 4 shows how mobility metrics significantly changed during social unrest in ECU and PAN and when a hurricane made landfall in MEX. In ECU, social protests[35] occurred between June 13 and June 30. $M_1$ dropped from approximately 20 km to 11.5 km, while $M_2$ raised from approximately 32% to 40%. In PAN, social protests[36] broke out on July 16 and lasted nearly two weeks. $M_1$ dropped from around 30 km to 17 km, while $M_2$ raised from approximately 27% to 32%. Between May 30 and 31, 2022, Hurricane Agatha hit the Oaxaca state of MEX with flooding and mudslides that killed at least 10 people and left 20 missing[37]. $M_1$ dropped from an average of around 28 km (on the previous days) to around 22 km. This drop is seen at the country level, so it was mitigated by the relatively small area (with respect to the area of MEX) impacted by the hurricane. $M_2$ did not change significantly with respect to the previous day.

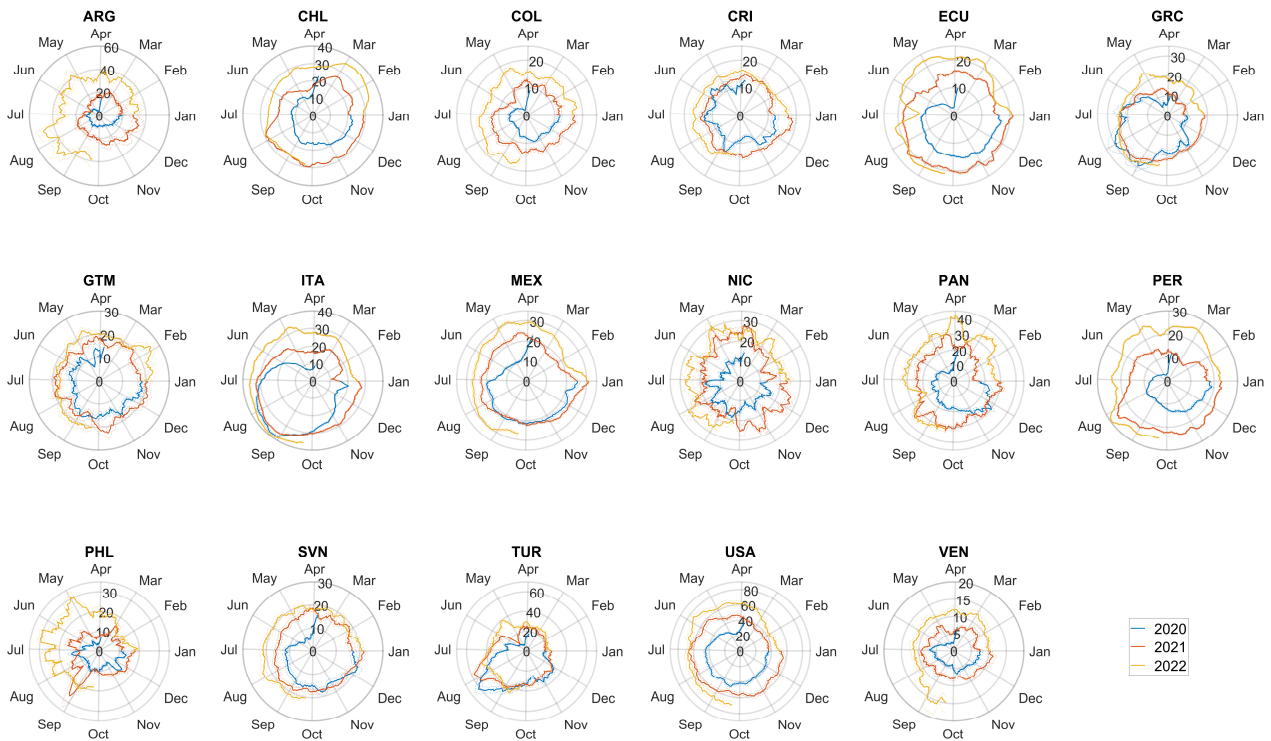

*Fig. 1 Estimated smoothed mobility metric $M_1$ (daily average travelled distance) based on a 14-day moving average from March 24, 2020, to September 22, 2022. The distance is expressed in kilometres. Dashed lines are 95% confidence bands.*

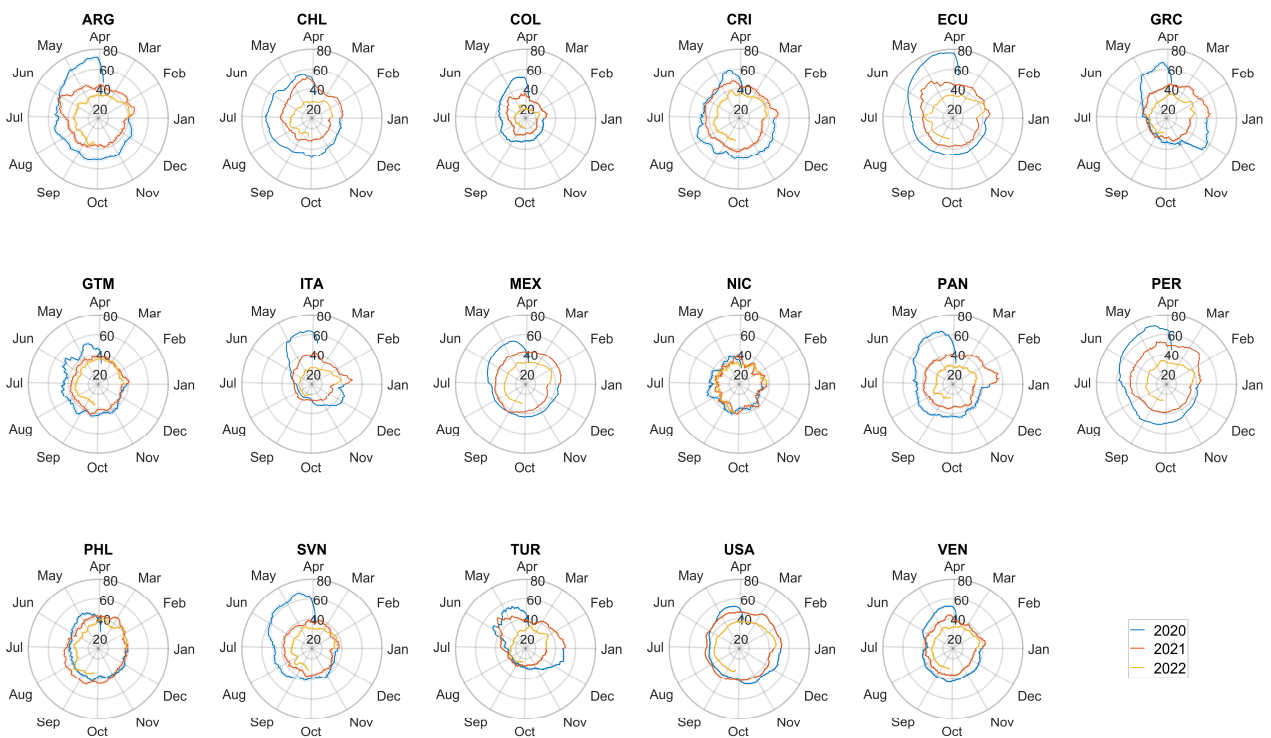

*Fig. 2 Estimated smoothed mobility metric $M_2$ (percentage of people who did not move during the 24 hours of the day) based on a 14-day moving average from March 24, 2020, to September 22, 2022. Dashed lines are 95% confidence bands.*

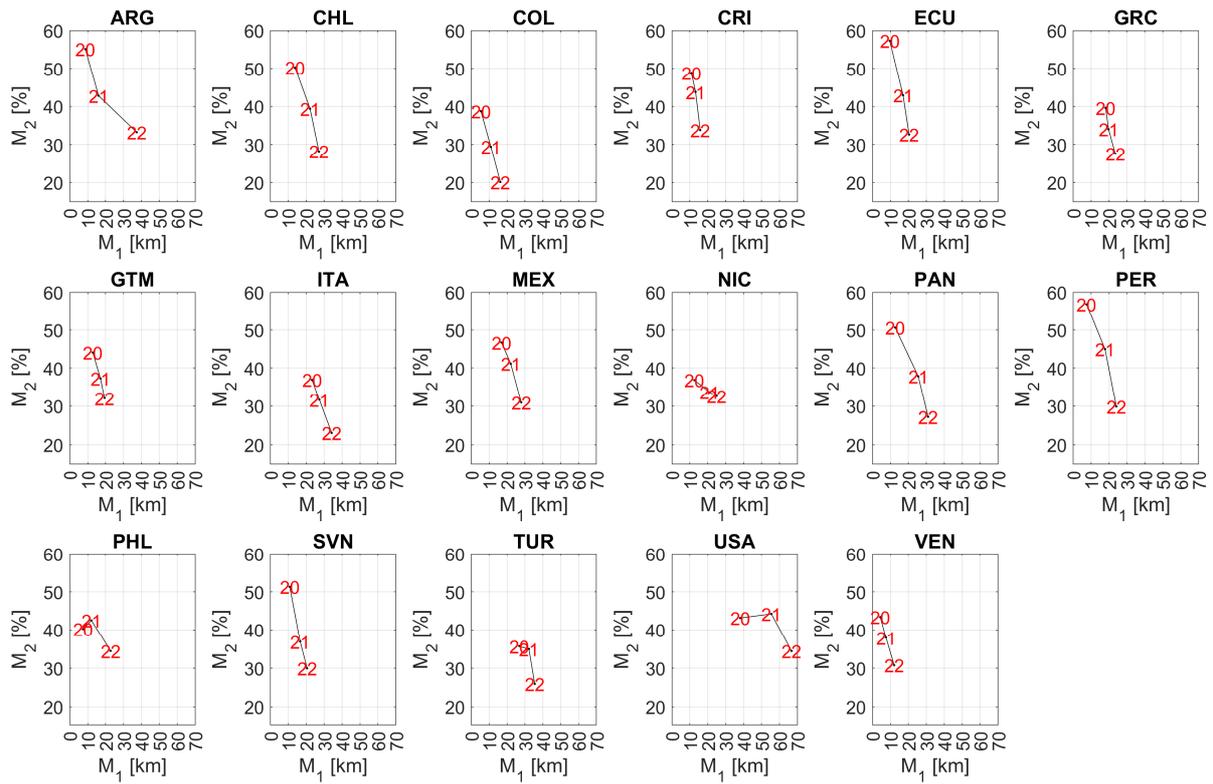

Fig. 3 $M_1$ and $M_2$ metrics averaged over the period from March 11 to September 22 in 2020, 2021 and 2022. In each plot, labels 20, 21 and 22 refer to years 2020, 2021 and 2022.

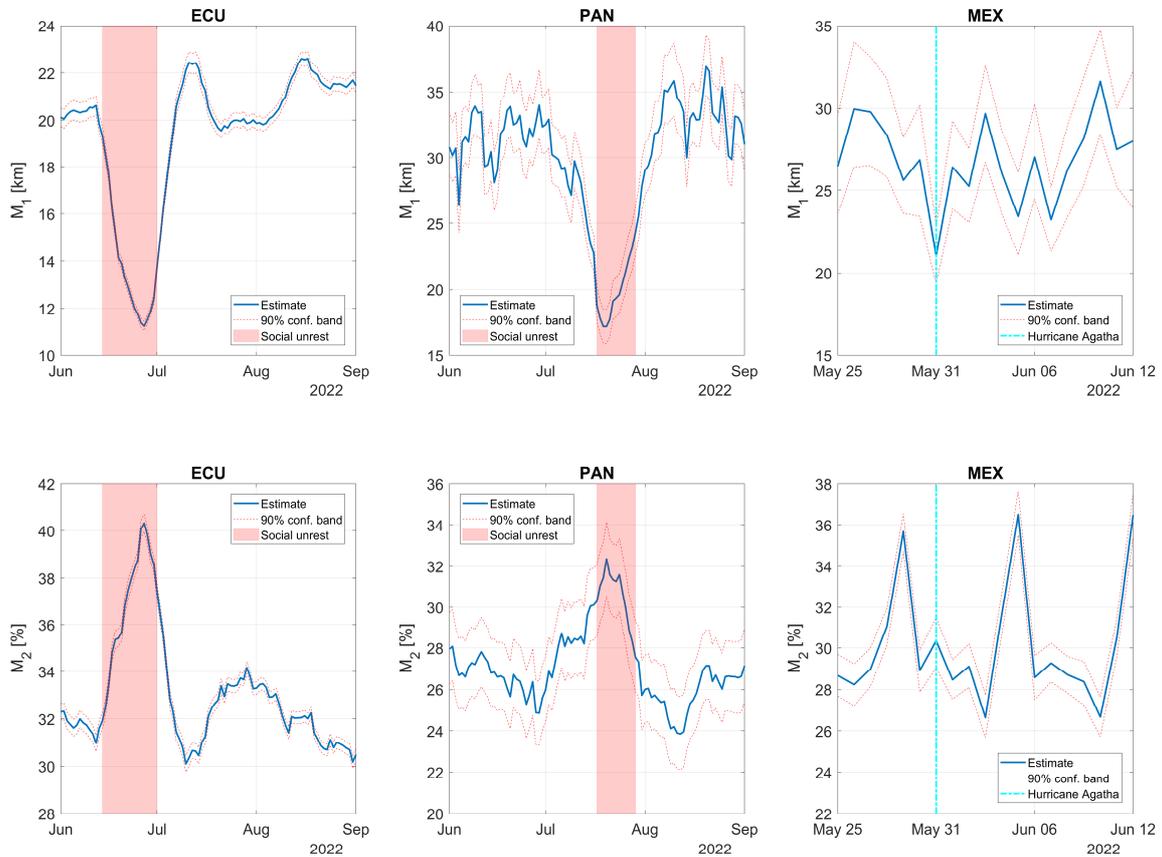

Fig. 4 Variations in mobility metrics $M_1$ and $M_2$ during 2022 social unrest in ECU and PAN and when Hurricane Agatha struck Oaxaca, MEX. For ECU and PAN, mobility metrics are based on 7-day smoothing. For MEX, no smoothing is applied.

## Comparison with Google's Community Mobility Reports

A comparison is made between metrics $M_1$ and $M_2$ and the mobility indices of the Google product. Google's mobility indices are given as variations (in percentage) in the number of visits to categories of places with respect to a baseline. This means that Google's indices do not carry exactly the same information of $M_1$ and $M_2$, nor the unit of measure is the same. Nonetheless, all indices are based on smartphone spatio-temporal trajectories, and their linear correlation is expected to be medium-high.

$M_1$ is compared with Google's "Transit stations" index and $M_2$ with Google's "Residential" index and with its "Workplaces" index. It is assumed that the "Transit stations" index positively correlates with the daily average distance travelled by people and that the "Residential" and the "Workplaces" indices correlate (positively and negatively, respectively) with the percentage of people who did not move during the 24 hours of the day.

Fig. 5 shows, for each country and for different levels of smoothing of the time series, the linear correlation (without sign) between the metrics and Google's indices. For most countries, correlations are high and significantly increase when moving from no smoothing to a 7-day moving average smoothing. The lowest correlations are exhibited by NIC, SVL and PHL, which are among the countries with the lowest average number of daily smartphone trajectories (average sample size) in the data set (see Fig. 6).

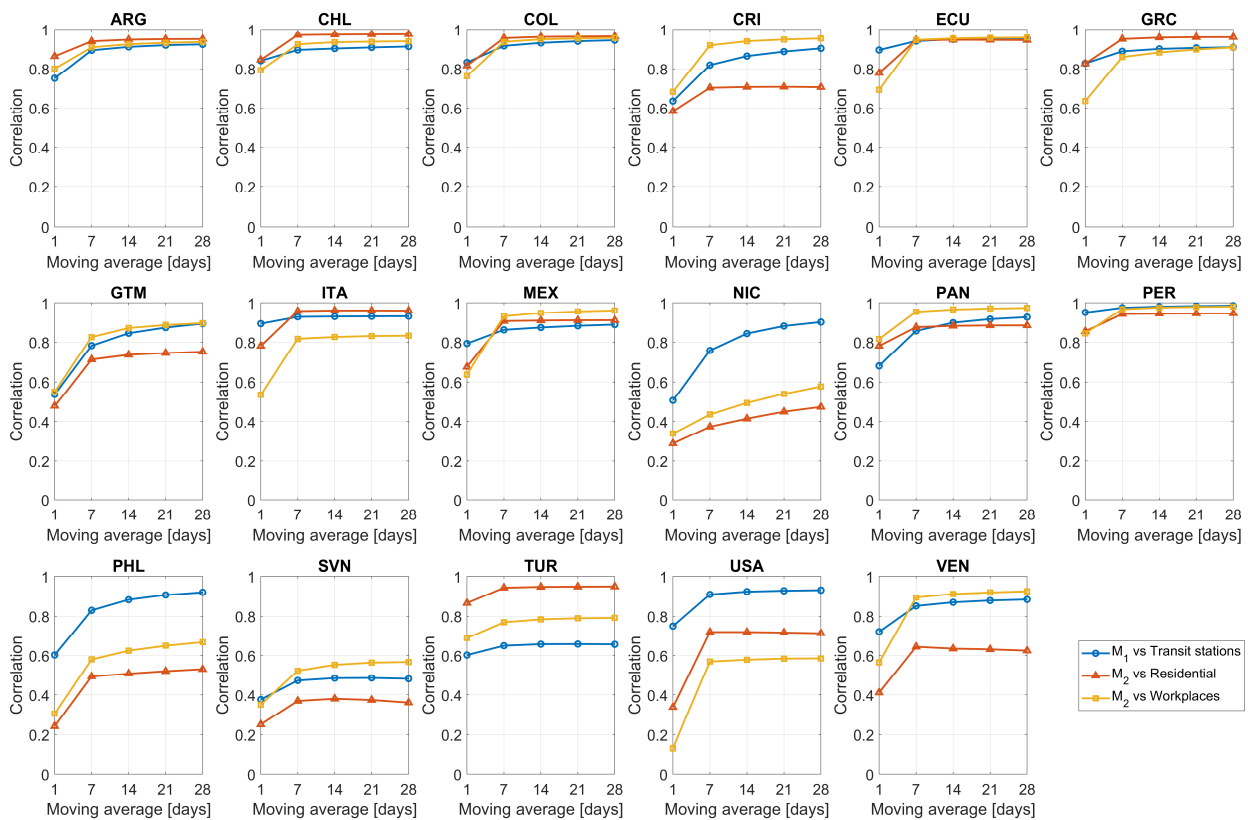

*Fig. 5 Linear correlation (without sign) between the $M_1$ mobility metric and Google's "Transit stations" mobility index, between the $M_2$ mobility metric and Google's "Residential" mobility index and between the $M_2$ mobility metric and Google's "Workplaces" mobility index; 1-day moving average means no smoothing.*

**Data set penetration analysis**

The relationship between correlations and average sample size is better described by accounting for country population size. The average sample size is computed for each country for the period from March 11, 2020, to September 22, 2022, and divided by country population size. This gives the average data set penetration in each country.

For each correlation between metrics and indices, and for both the non-smoothing and the 7-day moving average cases, the 17 average penetration values are related with the 17 correlations using a beta regression (see Methods section). Fig. 7 shows the data and modelling results. In general, the higher the average penetration, the higher the correlation between indices. The exception is given by the linear correlation between $M_1$ and the "Transit stations" index vs average penetration when the 7-day smoothing is applied (F-test p-value = 0.134). Additionally, average penetrations between $10^{-4}$ and $10^{-3}$ (i.e., 1 smartphone trajectory every 10,000 and every 1000 people, respectively) delimit the transition between medium (~0.5) to high (>0.7) correlations.

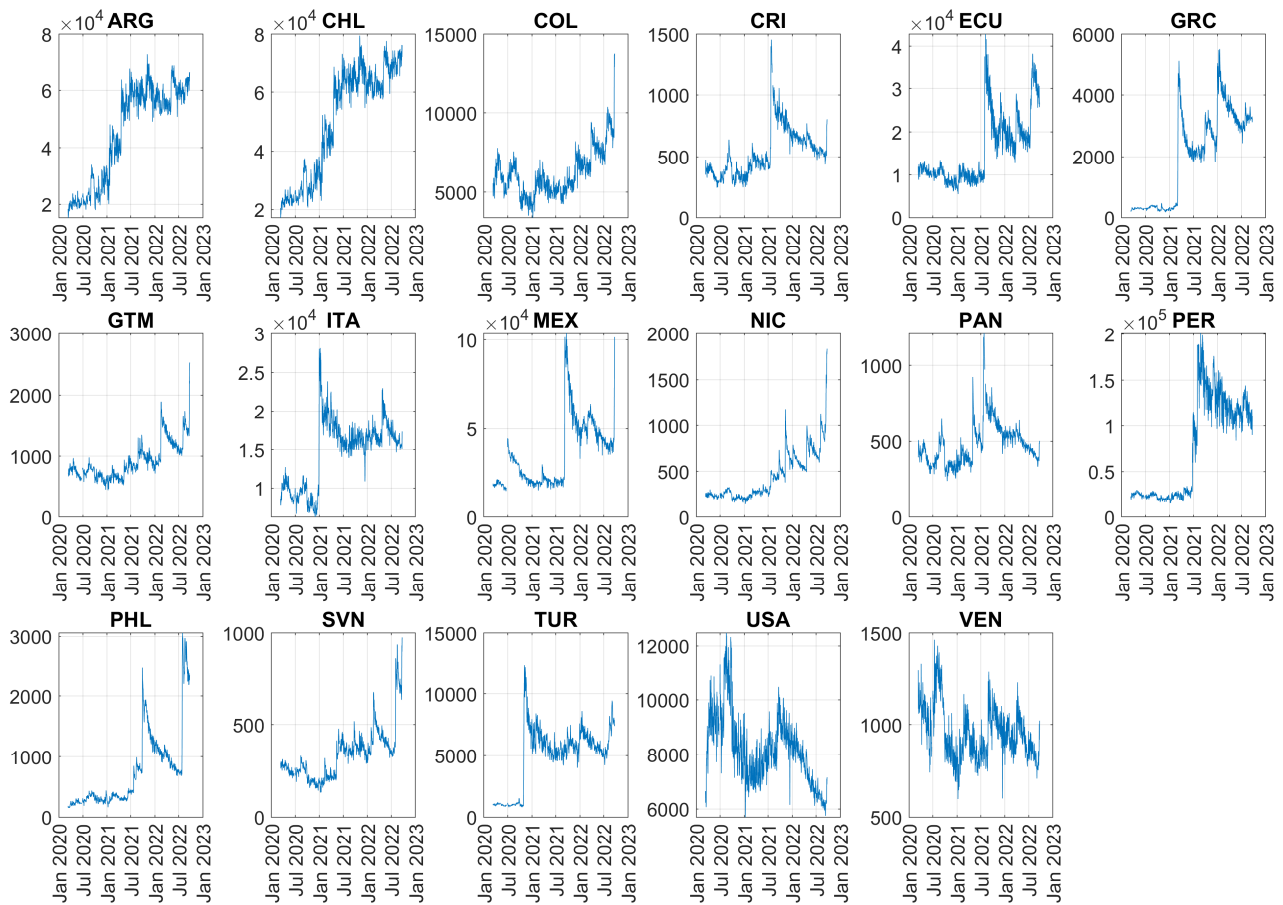

*Fig. 6 Time series of the number of daily smartphone trajectories used to estimate $M_1$ and $M_2$ metrics by country.*

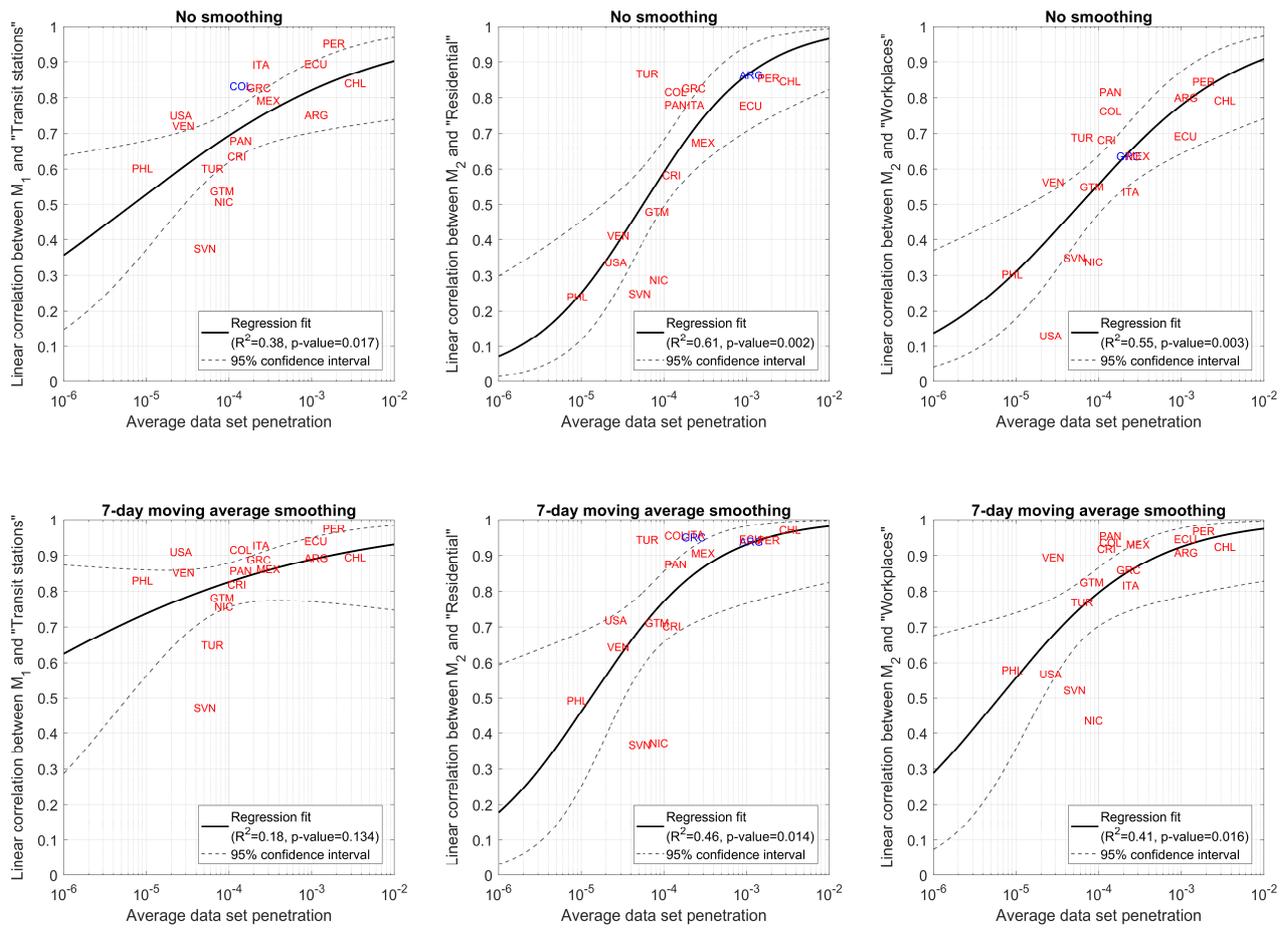

*Fig. 7 Linear correlation (without sign) between $M_1$ and $M_2$ and Google's mobility indices vs average data set penetration for the 17 countries under the non-smoothing and 7-day moving average smoothing cases. Solid lines are the fits by a beta regression model with logit link function, while dashed lines are 95% confidence intervals on fitted values. $R^2$ are pseudo coefficients of determination, while the p-values refer to an F-test on the regression model, which tests whether the model fits significantly better than a model with only the constant term. Each label is a country. Overlapping labels are in different colours.*

## Discussion

Mobility metrics and indices are proxies of societal health and stress. Their constant monitoring informs assessment of the impact of both long- and short-term adverse events. Thus, location data collected by smartphone apps play an important role, as they carry information on people mobility. With Apple's Mobility Trends Reports discontinued and Google's Community Mobility Reports no longer updated after October 2022, it is crucial to understand whether similar products can be produced from alternative mobility data sets and made available to the scientific community.

This study reconstructs smartphone trajectories from an independent mobility data set. These allow estimation of country-level mobility metrics with relatively high correlations to similar indices by Google. The average data set penetration among the country population requires 1 smartphone trajectory for every 10,000 people, which is a small value compared with that in Big Tech product coverage.

## Methods

### Trajectory description

The mobility data set analysed in this work includes smartphones trajectories collected from the 17 countries during the period from March 11, 2020, to September 22, 2022 (926 days). Each trajectory covers any possible subset of the 926 days (from only one day to the full period).

In general, the $k$th trajectory, $1 \leq k \leq K$, is composed of $M_k$ observations, with $K$ the total number of trajectories. The $m$th observation, $1 \leq m \leq M_k$, of the $k$th trajectory is given by

$$(ID_k, lat_m, lon_m, u_m, t_m), \qquad (1)$$

where $ID_k$ is the anonymized smartphone/trajectory identifier, $lat_m$ and $lon_m$ are the latitude and longitude smartphone coordinates, $u_m$ is the uncertainty on smartphone coordinates, and $t_m$ is the timestamp.

Uncertainty $u_m$ is in metres and represents the standard deviation (sigma) of two independent normal distributions centred on each smartphone coordinate. Timestamp $t_m$ is based on the country local time and refers to the time $lat_m$ and $lon_m$ observed by the smartphone. Timestamps have the following constraint: $min(t_{m+1} - t_m) \geq 20$ minutes.

### Trajectory sanitation

Trajectory observations which exhibit a latitude and longitude equal to zero are removed from the trajectory. Zero values are returned by a smartphone when geolocation is not possible. Also, $u_m$ values equal to zero or negative are replaced with an uncertainty equal to 25 metres. This is the typical uncertainty when a smartphone localizes itself using Wi-Fi networks and/or cell phone antennas. The percentage of trajectory observations affected by the replacement is 0.08%.

### Daily estimates of M₁ and M₂ metrics

M₁ and M₂ mobility metrics are estimated with daily temporal resolution. For each day $d = 1, \ldots, 926$, only trajectories with at least 12 observations between 00:00:00 and 23:59:59 local time and with a temporal span of at least 12 hours between the first and the last observation contribute to the estimate of M₁ and M₂. At the $d$th day and the $i$th country, the number of daily trajectories that satisfy the constraints above is denoted by $N_{d,i}$. In general, $N_{d,i} \neq N_{d',i}$ for $d \neq d'$.

For any two consecutive trajectory observations, the geodetic distance $l_{m,m+1}$ between $(lat_m, lon_m)$ and $(lat_{m+1}, lon_{m+1})$ is computed. The following transformation is then applied:

$$\hat{l}_{m,m+1} = \begin{cases} l_{m,m+1} & if \ l_{m,m+1} \geq u_m + u_{m+1} \\ 0 & otherwise \end{cases}. \qquad (2)$$

The transformation implies that the estimated travelled distance $\hat{l}_{m,m+1}$ between time $t_m$ and $t_{m+1}$ is greater than zero only if the 1-sigma uncertainty disks do not overlap. The daily travelled distance by the $k$th smartphone is given by:

$$\hat{L}_{d,i,k} = \sum_{m=1}^{M_k} \hat{l}_{m,m+1} I(t_m \in [\tau_d, \tau_{d+1})), \qquad (3)$$

where $I(t_m \in [\tau_d, \tau_{d+1}))$ is 1 if timestamp $t_m$ is within day $d$ and 0 otherwise (with $\tau_d$ the timestamp related with midnight of day $d$). Since $\hat{l}_{M_k, M_k+1}$ is computed using the last trajectory observation of day $d$ and the first of day $d+1$, Eq. ((3)) implies that any travel occurring across midnight contributes to the total travelled distance on day $d$.

Moreover, let

$$\hat{U}_{d,i,k} = \begin{cases} 1 & if\ \hat{L}_{d,i,k} < 0.2\ km \\ 0 & otherwise \end{cases} \qquad (4)$$

be a binary variable equal to 1 if the smartphone does not move during the 24 hours of the day and 0 otherwise. The threshold $0.2\ km$ is set to accommodate for indoor smartphone movements during the day that may sum up to a small distance.

For each country, first-level administrative divisions (regions) are considered. The daily average travelled distance for the $c$th region of the $i$th country is given by:

$$\hat{L}_{d,i,c} = \frac{1}{N_{d,i,c}} \sum_{k=1}^{N_{d,i}} \hat{L}_{d,i,k} \cdot I\left((\overline{lat}_k, \overline{lon}_k) \in R_c\right), \qquad (5)$$

where $I\left((\overline{lat}_k, \overline{lon}_k) \in R_c\right)$ is equal to 1 if the daily average smartphone coordinates $(\overline{lat}_k, \overline{lon}_k)$ are in the $R_c$ region, and 0 otherwise. $N_{d,i,c}$ is the number of daily trajectories in the $c$th region.

Let $p_{i,c}$ be the population count of the $c$th region, $1 \leq c \leq C_i$. The daily average distance for the $i$th country (mobility metric M₁) is given by

$$\hat{L}_{d,i} = \sum_{c=1}^{C_i} \hat{L}_{d,i,c} \cdot w_{i,c}, \qquad (6)$$

where

$$w_{i,c} = \frac{p_{i,c}}{\sum_{c=1}^{C_i} p_{i,c}}, \qquad (7)$$

is a weight based on the region population. The adoption of this weighting approach is dictated by three reasons: 1) the spatial distribution of smartphone-app users does not necessarily mimic the population distribution; 2) events affecting people mobility may be limited to some regions, or their strength vary across regions[38]; 3) in general, a weighting approach based on a population stratification helps reduce the bias of estimates[39].

By replacing $\hat{L}_{d,i,k}$ with $\hat{U}_{d,i,k}$ in Eq. ((5)) and following the same procedure described above, the mobility metric M$_2$ (i.e., $\hat{U}_{d,i}$) is computed for each day and country.

**Uncertainty assessment**

Uncertainty on daily M$_1$ and M$_2$ figures (i.e., $\hat{L}_{d,i}$ and $\hat{U}_{d,i}$, respectively) is assessed using a non-parametric bootstrap approach[40]. At the $b$th bootstrap iteration, $1 \leq b \leq B$, values $\hat{L}_{d,i,h,b}$ and $\hat{U}_{d,i,h,b}$, $1 \leq h \leq N_{d,i,c}$, are sampled with replacements from the observed $\hat{L}_{d,i,k}$ and $\hat{U}_{d,i,k}$ values restricted to the $c$th region. Following Eq. ((5))–((7)), the resampled values are used to produce the bootstrap sample $(\hat{L}_{d,i,1}, \ldots, \hat{L}_{d,i,B})$ and the bootstrap sample $(\hat{U}_{d,i,1}, \ldots, \hat{U}_{d,i,B})$. Fixing $B = 1000$, bootstrap samples are used to compute their empirical distribution. This allows evaluation of $(100 - \alpha)\%$ bootstrap confidence intervals[41] on the $\hat{L}_{d,i}$ and $\hat{U}_{d,i}$ estimates, with $\alpha$ equal to 5 in this work.

**Temporal smoothing**

Temporal smoothing of $\{\hat{L}_{d,i}\}$ and $\{\hat{U}_{d,i}\}$ time series is based on a $q$-day moving average, with $q$ equal to 7, 14, 21 and 28. The smoothed version of $\hat{L}_{d,i,c}$ is

$$\hat{L}^q_{d,i,c} = \frac{1}{\sum_{s=d-q+1}^{d} N_{s,i,c}} \sum_{s=d-q+1}^{d} \sum_{k=1}^{N_{s,i}} \hat{L}_{s,i,k} \cdot I\left((\overline{lat}_k, \overline{lon}_k) \in R_c\right). \qquad (8)$$

Similarly, $\hat{U}^q_{d,i,c}$ is defined by replacing $\hat{L}_{s,i,k}$ with $\hat{U}_{s,i,k}$ in Eq. ((8)). Confidence intervals on $\hat{L}^q_{d,i,c}$ and $\hat{U}^q_{d,i,c}$ are based on bootstrap samples which include $q$ days of resampled data. This allows obtaining confidence intervals with the correct width.

**Comparison with Google's Community Mobility Reports**

Community Mobility Reports by Google gives percentages of variation in the number of visits to place categories with respect to a baseline. The categories are "Retail and recreation", "Grocery and pharmacy", "Parks", "Transit stations", "Workplaces" and "Residential". For each category, time series of percentages of variation are available at both country and regional levels with daily temporal resolution.

M$_1$ and M$_2$ time series are compared with Google's country-level time series by computing linear correlations. M$_1$ is correlated with the "Transit stations" index and M$_2$ with both the "Workplaces" and "Residential" indices. Comparison is made using both non-smoothed and smoothed time series (i.e., $q \in \{7,14,21,28\}$). A highly positive or highly negative correlation means that M$_1$ and/or M$_2$ carry information on people mobility similar to that of Google's Community Mobility Reports.

**Beta regression on correlations vs average penetration**

A beta regression is adopted to describe the relationship between the data set average penetration and the correlations without sign $|\rho|$ between the non-smoothed M$_1$ and M$_2$ metrics and Google's indices. Beta regression is imposed by $|\rho| \in [0,1]$.

For the generic $i$th country, $|\rho_i| \sim \mathcal{B}(\mu_i, \phi)$, where $\phi$ is the precision parameter of the beta distribution and $g(\mu_i) = x_i'\boldsymbol{\beta}$, with $g$ the logit link function, $x_i$ the vector of regressors and $\boldsymbol{\beta}$ the vector of unknown model parameters. Here, $x_i = [1, log_{10}(\pi_i)]'$, with $\pi_i$ the average data set penetration for the $i$th country. Model fitting capability is described by the pseudo coefficient of determination $R^2 = corr(|\rho|, \widehat{|\rho|})^2$, with $\widehat{|\rho|}$ the model estimate. An F-test on the regression model is used to tests whether the model fits significantly better than a model with only the constant term (i.e., $x_i = 1$).

**Sensitivity analysis**

Estimates of M$_1$ and M$_2$ are based on three arbitrary choices. First, only trajectories with at least $n$ observations within a span of at least $n$ hours are used ($n = 12$). Second, $\hat{l}_{m,m+1} = l_{m,m+1}$ only if $l_{m,m+1} \geq ru_m + ru_{m+1}$ ($r = 1$, see Eq. ((2))). Third, $\widehat{U}_{d,i,k} = 1$ if $\hat{L}_{d,i,k} < z$ ($z = 0.2\ km$, see Eq. ((4)).

The choice of $n$ affects both M$_1$ and M$_2$, while the choice of $r$ and $z$ only affects M$_2$. Values used in this work are the result of a sensitivity analysis. Considering ITA and the period from March 11, 2020, to September 22, 2022, the correlation without sign $|\rho_1|$ between M$_1$ (non-smoothed) and the "Transit stations" index by Google and the correlation without sign $|\rho_2|$ between M$_2$ (non-smoothed) and the "Residential" index by Google are estimated for each combination of $n \in \{3,6,9,12,15\}$, $r \in \{1,2,3\}$ and $z \in \{0.1,0.2,0.3,0.4\}\ km$.

Considering all combinations, $|\rho_1|$ ranges between 0.878 and 0.893, while $|\rho_2|$ ranges between 0.766 and 0.781. Correlations are not significantly affected by large variations in $n$, $r$ and $z$. For both $|\rho_1|$ and $|\rho_2|$, the maximum is reached when $n = 12$, $r = 1$ and $z = 0.2\ km$.

**Data availability**

The MobMeter data set which includes the mobility metrics produced in this work is available on Zenodo at https://zenodo.org/record/6984638

Google's Community Mobility Reports are available at https://www.google.com/covid19/mobility/